\documentclass[a4paper,12pt]{article}

\usepackage{amsmath}
\usepackage{amsthm}
\usepackage{amsfonts}          % if you want the fonts
\usepackage{amssymb}           % if you want extra symbols

\newlength{\xtra}
\usepackage[all]{xy}           % Commutative diagrams

\newcommand{\eqdef}{\stackrel{\rm def}{=}}
\newcommand{\Q}{\ensuremath{{\mathbb{Q}}}}

\newcommand{\C}{\ensuremath{{\mathbb{C}}}}

\newcommand{\Z}{\mathbb{Z}}

\newcommand{\Tr}{\mathop{\rm Tr}\nolimits}

\newcommand{\Dbrane}{D--brane}

\newcommand{\Ktheory}{K--theory}

\newcommand{\Kgroups}{K--groups}

\newcommand{\Kunneth}{K{\"u}nneth}

\newcommand{\Tor}{{\mathop{\rm Tor}\nolimits}}
\newcommand{\Sym}{{\mathop{\rm Sym}\nolimits}}

\newcommand{\rank}{{\mathop{\rm rank}\nolimits}}
\newcommand{\codim}{{\mathop{\rm codim}\nolimits}}
\newcommand{\depth}{{\mathop{\rm depth}\nolimits}}
\newcommand{\lcm}{{\mathop{\rm lcm}\nolimits}}

\newcommand{\tKop}[1]{\ensuremath{\vphantom{K}^{#1}\!K}}
\newcommand{\tK}{\ensuremath{\tKop{t}}}
\newcommand{\ptset}{\ensuremath{\{\text{pt.}\}}}
\newcommand{\GTop}{{\ensuremath{G\text{-Top}}}}
\newcommand{\tGTop}{{\ensuremath{t\text{-}G\text{-Top}}}}

\newcommand{\textdef}[1]{\textbf{#1}}

{\end{list}}

{\everymath{\displaystyle\everymath{}}\array}%
{\endarray}

\newtheorem{theorem}{Theorem}

\newtheorem{definition}{Definition}

\begin{document}

\begin{titlepage}
  \begin{flushright}
    hep-th/0305178
    \\
    LPTENS-03/19    
  \end{flushright}
  \vspace*{\stretch{3}}
  \begin{center}
    \Huge \bf
    Twisted K--Theory of Lie Groups
  \end{center}
  \vspace*{\stretch{2}}
  \begin{center}
    {\large Volker Braun}\\
    \bigskip    
    \'Ecole Normale Sup\'erieure\\
    24 rue Lhomond\\
    75231 Paris, France\\
    \medskip
    \texttt{volker.braun@lpt.ens.fr}    
  \end{center}
  \vspace*{\stretch{1}}
  \begin{abstract}
    I determine the twisted \Ktheory{} of all compact simply connected
    simple Lie groups. The computation reduces via the
    Freed--Hopkins--Teleman theorem~\cite{FreedVerlindeAlg} to the CFT
    prescription, and thus explains why it gives the correct result.
    Finally I analyze the exceptions noted by Bouwknegt et
    al~\cite{Bouwknegt}.
  \end{abstract}
  % \keywords{Conformal and W Symmetry, D-branes}  
  \vspace*{\stretch{5}}
\end{titlepage}

\newpage

\pagestyle{headings}
\tableofcontents

\section{Introduction}
\label{sec:intro}

WZW models provide a nice playground of solvable conformal field
theories. The key to their treatment is of course that the target
space is a group manifold $G$, and everything in the theory should be
expressed in terms of the representation theory of this group. 

An example for this is the fusion ring of the CFT, which is determined
by the representation theory of the loop group $LG$ at the
corresponding level.  A more sophisticated instance are the Cardy
branes, special boundary states in the CFT labeled by the irreducible
representations.

In~\cite{AlekseevSchomerus,FredenhagenSchomerus} this was combined
with boundary RG flow methods to determine possible decay paths of
boundary states. The authors then used this to guess the underlying
conserved charges and compared it with the expected result, the
twisted \Ktheory{} of $G$.

This check was significantly improved in~\cite{MaldacenaMooreSeiberg},
where the predicted charge groups for all $G=SU(n)$ are compared with
a purely K--theoretic computation. However the \Ktheory{} computation
uses an explicit cell decomposition and thus neither generalizes
to other Lie groups nor does it make use of the representation
theory. So although one finds at the end that the charge groups match
this is not a very satisfactory result.

My work closes this gap: I reduce the computation of the twisted
\Ktheory{} to a calculation in representation theory, and the latter
will be very similar to the CFT calculation. So there will be no
mystery that the resulting charge groups match, and furthermore it
works for all simply connected, compact, simple Lie groups in the same way.
The crucial connection between CFT and \Ktheory{} is the theorem of
Freed, Hopkins and Teleman~\cite{FreedVerlindeAlg}.

The actual computation is closely related to the method
in~\cite{SakuraDraft} and I would like to thank Sakura
Sch\"afer--Nameki for sharing her draft with me. However the seemingly
simpler case of a WZW model instead of a coset is technically quite
involved, and this is what I will be dealing with.

During the preparation of this paper a comment of Hopkins was relayed
through~\cite{MooreTalk}. Although it suffered from the Chinese
whispers phenomenon it clearly suggested to compute the twisted
\Ktheory{} along the following lines.

\section{An Example}
\label{sec:example}

Take the simplest case, the Lie group $G=SU(2)\simeq S^3$. The WZW
model at level $k$ contains Cardy boundary states
$\big|\lambda\big>\!\big>$, labeled by the first $k+1$ irreps of
$SU(2)$ (or alternatively by irreducible positive energy irreps of
$\smash{\widetilde{LG}}$). Call this index set $\mathcal{J}_k$. Now it
was argued in~\cite{FredenhagenSchomerus} that the charges $q_\lambda$
of the boundary states satisfy
\begin{equation}
  \label{eq:RG}
  \dim_\C(\mu)\, q_\lambda = 
  \sum_{\nu\in \mathcal{J}_k} N_{\mu \lambda}^\nu \, q_\nu
\end{equation}
where $N_{\mu \lambda}^\nu$ are the fusion coefficients. This has two
immediate consequences:
\begin{itemize}
\item Take $\lambda=0$ the trivial representation
  $\Rightarrow~q_\mu = \dim_\C(\mu)\, q_0$, every charge is a multiple
  of $q_0$.
\item Any relation $\sum_i a_i \mu_i=0$ in the fusion ring leads to
  \begin{equation}
    \left( 
      \sum_{i\in \mathcal{J}_k} 
      a_i \dim_\C(\mu_i) 
    \right) 
     \, q_0 = 0
  \end{equation}
  so the charge $q_0$ is torsion. If there are no further
  identifications than eq.~\eqref{eq:RG} then the order of the torsion
  is the minimal dimension of a relation in the fusion ring.
\end{itemize}
Especially for $SU(2)$ at level $k$ the relations in the fusion ideal
have dimensions $(k+2)\Z$, therefore the charge group is $\Z_{k+2}$.

Let us compare this with the twisted \Ktheory{} of $SU(2)$. It is
readily evaluated for any given twist class $\tau \in
H^3\big(SU(2)\big)\simeq \Z$, for example Rosenberg's spectral
sequence~\cite{RosenbergSS} yields immediately that
\begin{equation}
\label{eq:twistedKresult}
  \tKop{\tau}^\ast\big(SU(2)\big) = \Z_\tau
\end{equation}
Although the result nicely matches the CFT if you identify $\tau =
k+2$ it does not explain why this should be so. Even from the
physicist's perspective this is something of a miracle since the WZW
model only has to reproduce the classical geometry in the $k\to
\infty$ limit.

I will now explain why the CFT formula works, details of the
computation will be explained in the following sections. First
introduce $G$ equivariant \Ktheory{} via
\begin{equation}
\label{eq:tKrewrite}
  \tKop{\tau}^\ast\big(G\big) =
  \tKop{\tau}^\ast_{G}\Big(
    G_\text{Tr}\times G_\text{L}  \Big) = 
  \tKop{\tau}^\ast_{G}\Big(
    G_\text{Ad}\times G_\text{L}\Big) 
\end{equation}
where the subscripts $\text{Tr}$, $\text{L}$, $\text{Ad}$ denote the
group action on that factor as \underline{Tr}ivial action,
\underline{L}eft multiplication, \underline{Ad}joint action. Now the
twisted \Ktheory{} for a Cartesian product can be computed via a
\Kunneth{} spectral sequence, the result is that
\begin{equation}
\begin{split}
\label{eq:tKtor}
  \tKop{\tau}^\ast_{G}\Big(
    G_\text{Ad}\times G_\text{L}\Big) 
  \,=&\, 
  \Tor_{K_G(\ptset)}^\ast \Big(   
      \tKop{\tau}^\ast_{G}\big(G_\text{Ad}\big),\, 
      K^\ast_{G}\big(G_\text{L} \big)
    \Big)     
\\ \,=&\,
  \Tor_{RG}^\ast \big( RG/I_k,\, \Z \big)
\end{split}
\end{equation}
with the Verlinde algebra\footnote{\textdef{Verlinde algebra} will
  always denote the algebra over $\Z$, i.e.
  $RG\simeq\Z\big[x_1,\dots,x_{\rank (G)}\big]$ modulo the ideal
  generated by the fusion rules. The associated algebra over $\Q$ will
  be called \textdef{rational Verlinde algebra}, similarly
  complexified Verlinde algebra. In the language of
  \cite{FuchsReview}, the Verlinde algebra is the fusion ring and the
  complexified Verlinde algebra is the fusion rule algebra.}  $RG/I_k$
at level $k=\tau-\check{h}(G)$, where $\check{h}(G)$ is the dual
Coxeter number. Especially for $G=SU(2)$ the representation ring is
\begin{equation}
  RSU(2) = \Z[\Lambda]
\end{equation}
where $\Lambda$ is the fundamental $2$-dimensional representation. In
this case the Verlinde algebra is particularly easy to write down. The
ideal of relations is generated by $(k+1)$--th symmetric power of
$\Lambda$, i.e. the irreducible representation in dimension $k+2$. One
can get this representation by reducing the $(k+1)$--fold tensor
product of $\Lambda$:
\begin{equation}
  I_k =  
  \left< \Sym_{k+1} \Lambda \right> = 
  \left< \Lambda^{k+1}-\text{lower order terms} \right>,
\end{equation}
for example $\Sym_2\Lambda = \Lambda^2-1$. 
Now evaluate eq.~\eqref{eq:tKtor}. $\Tor$ is the derived functor of
$\otimes$ which means that we have to do the following:
\begin{enumerate}
\item Find a projective resolution of the Verlinde algebra, i.e. a
  complex whose homology is $RG/I_k$ at position $0$ and all entries
  projective.
\item Apply $-\otimes_{RG}\Z$.
\item Compute the homology of the resulting complex. 
\end{enumerate}
Here is a (free hence) projective resolution:
\begin{equation}  
  \label{eq:simplekoszul}
  P^\bullet \eqdef \quad
  \xymatrix{
    0 \ar[r] &
    \Z[\Lambda] \ar[rr]^{\cdot \big(\Sym_{k+1}\Lambda\big)} &&
    \underline{\Z[\Lambda]} \ar[r] &
    0
  }
\end{equation}
where the underlined entry is at position $0$. For any ring
$R\otimes_R \Z=\Z$, the only problem is to find the map induced from
multiplying with $\Sym_{k+1}\Lambda$. For that we have to remember
that the $\Z[\Lambda]$--module structure on $\Z$ comes from the
$K_G(\ptset)$--module structure on $K(\ptset)$, i.e. tensoring a
vectorspace with a representation. Since $K(\ptset)\simeq \Z$ is the
dimension of the vector space this is just multiplication with the
dimension of the representation. In our case
$\dim_{\C{}}\big(\Sym_{k+1}\Lambda\big) = k+2$, so
\begin{equation}
\label{eq:TorApplied}
  P^\bullet \otimes_{\Z[\Lambda]} \Z = \quad
  \xymatrix{
    0 \ar[r] &
    \Z \ar[r]^{k+2} &
    \underline{\Z} \ar[r] &
    0
  }  
\end{equation}
which is exact except in position $0$ where the homology is
$\Z_{k+2}$.

We have found that
\begin{equation}
  \tKop{k+\check{h}}^\ast\big(SU(2)\big) =
  \tKop{k+2}^\ast\big(SU(2)\big) =
  \Tor_{RSU(2)}^\ast \big( RSU(2)/I_k,\, \Z \big) = 
  \Z_{k+2}
\end{equation}
as we already know from eq.~\eqref{eq:twistedKresult}. But now it is
clear from eq.~\eqref{eq:TorApplied} that we got it as $\Z/\dim(I_k)$,
which is precisely what the CFT predicted. Furthermore
eqns.~\eqref{eq:tKrewrite},\eqref{eq:tKtor} hold for all simply
connected, compact, simple Lie groups, and I will use it to determine
all \Kgroups{} in the following.

\section{The general computation}
\label{sec:general}

Fix once and for all a simply connected compact simple Lie group $G$.
In the following we will work with topological spaces, together with
group actions on them and twist classes on them (the category
$\tGTop$). To facilitate this I will use the following conventions:

For a given $X\in \mathop{Ob}(t\text{-}G\text{-Top})$ let
$X^\GTop$ be the underlying $G$-space and $t_X\in
H^3_G\big(X^\GTop \big)$ the twist class. A map $h:X\to Y$ is
a $G$-map $h^\GTop: X^\GTop\to Y^\GTop$ such that $t_Y=h^\ast
\big( t_X \big)$; In general there need neither be a map from nor to a
point.

The Cartesian product $X\times Y$ is the ordinary product with twist
class $t_{X\times Y} = \pi_1^\ast (t_X) + \pi_2^\ast (t_Y)$, where
$\pi_{1,2}$ is the projection (they are not maps in \tGTop) to the
first and second factor.

The $t_X$-twisted equivariant \Ktheory{} will be denoted $\tK_G(X)$.

\subsection{The action map}
\label{sec:actionmap}

Write $G_\text{L}$ for $G$ acting on itself by left multiplication and
no twist, $t_{G_L}=0$. Furthermore let $G_\text{Tr}$ be $G$ acting
trivially on itself, but with arbitrary twist class $t_{G_\text{Tr}}
\in H^3_G(G_\text{Tr})\simeq \Z$.

The all-important observation is the following~\cite{SakuraDraft}:
For $G_\text{Ad}$ ($G$ with the adjoint
action on itself) one can pick a twist class such that there is an
isomorphism 
\begin{equation}
  \label{eq:flip}
  f:\, 
  G_\text{Ad} \times G_\text{L}\, \stackrel{\sim}{\to} \,
  G_\text{Tr} \times G_\text{L}
\end{equation}
This is the map I used previously in eq.~\eqref{eq:tKrewrite}.

On the underlying $G$-spaces, $f$ is of course the $G$-diffeomorphism
\begin{equation}
  \label{eq:flipGTop}
  f^\GTop:\, 
  G_\text{Ad}^\GTop \times G_\text{L}^\GTop\, \stackrel{\sim}{\to} \,
  G_\text{Tr}^\GTop \times G_\text{L}^\GTop
  ,~
  (g,\gamma)\mapsto (\gamma^{-1} g \gamma, \gamma)
\end{equation}
and since it identifies the cohomology groups we can pick a suitable
twist class on $G_\text{Ad}^\GTop \times G_\text{L}^\GTop$, it remains
to show that it comes from the projection on the first factor.

So (dropping the superscripts \GTop{} for readability) consider 
\begin{equation}
  p: G_\text{Ad} \times G_\text{L} \to G_\text{Ad}
\end{equation}
We want to show that $p^\ast: H^3_G\big(G_\text{Ad}\big) \to
H^3_G\big( G_\text{Ad} \times G_\text{L} \big) \simeq \Z$ is an
isomorphism. By the universal coefficient theorem and the Hurewicz
isomorphism it suffices to show that 
\begin{equation}
  p_*: \pi_i
  \overbrace{
    \Big( \big(G_\text{Ad} \times G_\text{L} 
                           \times EG)/G \Big)
  }^{\simeq G_\text{Ad}}
 \to \pi_i\Big( \big( G_\text{Ad} \times EG\big) /G \Big)
\end{equation}
is an isomorphism for $i=2,3$. This follows from the long exact
homotopy sequence of the fibration $p$:
\begin{equation}
  \pi_i(G) \to 
  \pi_i(G_\text{Ad})
  \stackrel{p_\ast}{\to}
  \pi_i\Big( \big( G_\text{Ad} \times EG\big) /G \Big)
  \to \overbrace{\pi_{i-1}(G)}^{=0~\text{for}~i=2,3} \to \cdots
\end{equation}
Following the image of the generating $S^3\subset G$ one sees that the
image of the leftmost map is contractible (zero) since the $G$-action
on $G_\text{Ad}$ had fixed points.

Obviously isomorphic spaces have the same cohomology, thus
\begin{equation}
  \label{eq:Kactionmap}
  \tK^\ast\big(G\big) =
  \tK^\ast_{G}\Big(
    G_\text{Tr}\times G_\text{L}  \Big) = 
  \tK^\ast_{G}\Big(
    G_\text{Ad}\times G_\text{L}\Big)   
\end{equation}

\subsection{Making contact with FHT}
\label{sec:rewriteFHT}

In the next section I will discuss how the \Ktheory{} of the factors
determines the \Ktheory{} of $G_\text{Ad}\times G_\text{L}$, for now I
will summarize the \Ktheory{} of the factors, and their $RG$--module
structure (i.e. how tensoring with a ordinary $G$--representation
acts). The easy part is 
\begin{equation}
  K^i_G(G_\text{L}) = K^i\big(\ptset\big) = 
  \begin{cases}
    0 & i~\text{odd} \\
    \Z & i~\text{even}
  \end{cases}
\end{equation}
with $RG$--action
\begin{equation}
  \rho \cdot x = \dim_\C(\rho) x 
  \qquad \forall \rho \in RG, x\in K^\ast_G(G_\text{L})\simeq \Z
\end{equation}
The twisted equivariant \Ktheory{} of $G_\text{Ad}$ is much more
complicated and given by
\begin{theorem}[Freed--Hopkins--Teleman=FHT]
  \begin{equation}
    \label{eq:FHT}
    \tK^{\dim G}_G(G_\text{Ad}) = RG / I_k  
  \end{equation}
  is an isomorphism of $RG$--modules, where the level
  $k=t_{G_\text{Ad}}-\check{h}$. 
\end{theorem}

\subsection{The \Kunneth{} Spectral Sequence}
\label{sec:kunneth}

A \Kunneth{} theorem in general computes the cohomology of a Cartesian
product $X\times Y$ from the cohomology of $X$ and $Y$. Now if you are
really lucky you find that $h^\ast(X\times Y) = h^\ast(X) \otimes
h^\ast(Y)$, for example if $h^\ast$ is the usual de Rham
cohomology. But in general this is far more difficult.

A rather exhaustive account for untwisted equivariant \Ktheory{} is
given in~\cite{Hodgkin}. One finds a \Kunneth{} spectral sequence
which --- to make things even worse --- does not always yield the
\Ktheory{} of the product. However for sufficiently nice groups (like
$G$ compact simply connected Lie groups, as is the case we are
interested in) this spectral sequence does compute the desired
\Ktheory{} of the product. Unfortunately the proof in~\cite{Hodgkin}
uses knowledge of the non-equivariant \Ktheory{} of
$G$, which is what we are after in the twisted case.

Fortunately the \Kunneth{} spectral sequence was later extended
in~\cite{RosenbergSchochet} to \Ktheory{} for $C^\ast$--algebras. This
is useful since we can think of twisted \Ktheory{} as the \Ktheory{}
of some (possibly noncommutative) $C^\ast$--algebra,
see~\cite{BouwknegtMathai}, and we get the following
\begin{theorem}
  \label{thm:twistedKunneth}
  Let $G$ be a simply connected compact Lie group, $X\in
  \mathop{Ob}(\tGTop)$ and $Y \in \mathop{Ob}(\GTop)$ (i.e. there is
  no twist on $Y$). Then there is a spectral sequence with
  \begin{equation}
    E_2^{-p,\ast} =  \Tor_{RG}^p \Big(   
      \tK^\ast_{G}\big(X\big), \, 
      K^\ast_{G}\big(Y\big)
    \Big)     
  \end{equation} 
  converging to $\tK_G(X\times Y)$.
\end{theorem}
In section~\ref{sec:tor} I will determine the $E_2$ term for the case
at hand. Of course we then have to worry about higher differentials
and extension ambiguities, but we will see in
section~\ref{sect:multiplication} that 
\begin{equation}
  \label{eq:tKtorResult}
  \tK^\ast(G_\text{Ad}\times G_\text{L})\simeq 
  E_2^{\ast,\ast} = \Tor^\ast(RG/I_k, \Z).  
\end{equation}

\subsection{Properties of the Verlinde algebra}
\label{sec:verlindeprop}

Of course it is possible to work out the Verlinde algebra for any
given group and level. However to actually compute the $\Tor$ in
general it would be very helpful if $RG/I_k$ were a complete
intersection, that is $I_k$ generated by a regular sequence (this will
be defined in the following).

Although it seems to be true, there is no general proof so far. I will
show that it would be a consequence of the existence of fusion
potentials.

\subsubsection{Fusion potentials}
\label{sec:fusionpotential}

A fusion potential is a polynomial 
\begin{equation}
  \phi\in RG\otimes\Q = \Q\big[ x_1,
  \dots, x_n \big]
  \,, \qquad 
  n=\rank\,G 
\end{equation}
such that the Verlinde ideal
$I_k\subset RG$ is generated by the partial derivatives:
\begin{equation}
  I_k = \left< 
    \frac{\partial \phi}{\partial x_1}, \dots, 
    \frac{\partial \phi}{\partial x_n}
  \right>_{RG}
\end{equation}
Such potentials have been determined for the $A_n=SU(n+1)$ and
$C_n=Sp(2n)$ Lie groups for all $k$. Gepner~\cite{GepnerGeom}
conjectured that such a potential exists in general, but I do not know
of any proof so far. In the following I will assume this to be true.

Actually I will only be using that the Verlinde ideal is generated by
$n$ elements. I do not know any proof for this weaker statement either.

\subsubsection{Regular sequences}
\label{sec:regularseq}

First let us recall the following
\begin{definition}
  Let $R$ be a ring, then an ordered sequence of elements
  $y_1,y_2,\dots,y_n\in R$ is a \textdef{regular sequence} on R if
  \begin{enumerate}
  \item They do not generate the whole ring:
    $\left<y_1,\dots,y_n\right> \not= R$
  \item For $i=1,\dots,n$, $y_i$ is a nonzerodivisor in
    $R/\left<y_1,\dots,y_{i-1}\right>$.
  \end{enumerate} 
\end{definition}
In general an ideal will have different sets of generators, some might
form a regular sequence and some will not. So to make everything
explicit (for $R=RG$) one would have to write down generators of the
Verlinde ideal for each Lie group $G$ and each level $k$, and then
show that the chosen generators in the chosen ordering form a regular
sequence.

Instead, I will take a more high--powered approach and only show the
existence of a regular sequence which generates the Verlinde
ideal. This is a rather elementary application of the theory of
commutative rings. 

First we note that the representation ring of $G$ is a polynomial ring
generated by the $n\eqdef \rank(G)$ fundamental representations, $RG
\simeq \Z\big[ x_1, \dots, x_n \big]$. Since it is a polynomial ring
over $\Z$ it is Cohen--Macaulay. This means that for each ideal $I\in
RG$ we have\footnote{Here (Co)Dimension always refers to the Krull
  dimension, i.e. lengths of chains of prime ideals. The
  Cohen--Macaulay property means that this definition retains some of
  the intuitive properties of ``dimension''.} $\codim(I) = \depth
(I)$. The same holds for the rational representation ring $RG\otimes
\Q$.

To simplify notation I will use a subscript $\Q$ to denote the change
of base ring from $\Z$ to $\Q$, so if $RG/I$ is the Verlinde algebra
then $RG_\Q/I_\Q$ is the rational Verlinde algebra (I will suppress
the level to avoid double subscripts).

The second ingredient is the fact that we are dealing with a RCFT, so
the Verlinde algebra is a finite rank torsion free $\Z$--module, i.e.
$RG/I \simeq \Z^d$ for some $d\in \Z_>$.

Especially $\dim\big( RG_\Q/I_\Q \big)=0$: From the point of view of
algebraic geometry, the spectrum of the rational Verlinde algebra is
just a finite set of points. Put differently, the minimal primes over
$I_\Q$ are maximal ideals.

Recall that the codimension of $I_\Q$ is the minimum over all prime
ideals $\mathfrak{p}\supset I_\Q$ over the maximum of lengths of
chains of prime ideals descending from $\mathfrak{p}$. If
$\mathfrak{p}$ is maximal then those chains have maximal length and we
get $\codim\, I_\Q = \dim\, RG_\Q = n$, the rank of $G$.

Now we actually want the codimension of $I$, and not its rational
version $I_\Q$. A fancy way of going from $\Z$ to the quotient field
$\Q$ is localization at the nonzero integers. The advantage of this
description is that it yields a nice relation between the prime ideals
in the corresponding polynomial rings, and we get (see e.g.
Theorem~36 of~\cite{Kaplansky})
\begin{equation}
  \codim~I = \codim~I_\Q = n
\end{equation}
The last piece of information we need is the only point specific to
WZW models: the Verlinde ideal $I$ can be generated by $n$ elements.

By the above remarks we know that $n=\depth\, I$, so we can apply
Theorem~125 of~\cite{Kaplansky}: If an ideal $I$ can be generated by
$n=\depth\, I$ elements then it can be generated by a regular sequence
(of $n$ elements).
\begin{theorem}
  There exist a regular sequence $y_1,\dots, y_n\in RG$ of length
  $n=\rank(G)$ that generates the Verlinde ideal: $\left<
    y_1,\dots,y_n \right> = I_k$.
\end{theorem}

\subsubsection{Koszul resolutions}
\label{sec:koszul}

Given any sequence $y_1,\dots,y_n\in RG$ the \textdef{Koszul complex}
$\mathcal{K}(y_1,\dots,y_n)$ is a complex of length $n+1$ with the
$i^{\text{th}}$ entry the degree $i$ piece of the exterior algebra
$\bigwedge_n RG$ (see~\cite{Eisenbud} for a nice introduction).
Another useful way of thinking about the Koszul complex would be the
following: First for one element, $\mathcal{K}(y_1)$ is the length two
complex analogous to the one depicted in eq.~\eqref{eq:simplekoszul}.
Then $\mathcal{K}(y_1,\dots,y_{i+1}) = \mathcal{K}(y_1,\dots,y_i)
\otimes \mathcal{K}(y_n)$ (of course now most subtleties are hidden in
the tensor product of complexes).

The whole point of this construction is that if $y_1,\dots,y_n$ form a
regular sequence in $RG$, then $\mathcal{K}(y_1,\dots,y_n)$ is a
resolution of the quotient ring $RG/\left<y_1,\dots,y_n\right>$.
Especially we can choose (by the preceding section) a regular sequence
generating the Verlinde ideal and thus obtain a resolution of the
Verlinde algebra.

\subsection{Deriving Tensor}
\label{sec:tor}

Finally we have everything to compute the $E_2$ term of the spectral
sequence in theorem~\ref{thm:twistedKunneth}. We know a projective
resolution of the Verlinde algebra, and now it is a rather simple
algebraic task to compute the $\Tor$.

Using the Koszul resolution we readily compute 
\begin{equation}
  \begin{split}
    \Tor_{RG}^\ast\Big( RG/I_k, ~\Z \Big) &= 
    \Tor_{RG}^\ast\Big( RG/\left<y_1,\dots,y_n\right>, ~\Z \Big) = 
    \\ &=
    H_\ast\Big( \mathcal{K}(y_1, \dots y_n) \otimes_{RG} \Z \Big)
    \\ &=
    H_\ast\left( \Big( \bigotimes_{i=1}^n \mathcal{K}(y_i) \Big)
      \otimes_{RG} \Z \right)
    \\ &=
    H_\ast\left( \bigotimes_{i=1}^n \big( \mathcal{K}(y_i) 
      \otimes_{RG} \Z \big) 
    \right)
    \\ &=
    H_\ast\left( \bigotimes_{i=1}^n 
      \big[ 0\to \Z \stackrel{d_i}{\longrightarrow}
      \underline{\Z}\to 0 \big] 
    \right)
  \end{split}
\end{equation}
where multiplication by $d_i = \dim(y_i)$ is the map induced by $y_i$
as explained in section~\ref{sec:example}. 

The homology of each factor is of course
\begin{equation}
    H_r\left( \big[ 0\to \Z \stackrel{d_i}{\longrightarrow}
      \underline{\Z}\to 0 \big] 
    \right) = 
    \begin{cases}
      \Z_{d_i} & r=0 \\ 0 & \text{else}
    \end{cases},
\end{equation}
and all that remains is to apply the usual K\"unneth formula for chain
complexes, the result is 
\begin{equation}
  \label{eq:torresult}
  H\left( \bigotimes_{i=1}^n
    \big[ 0\to \Z \stackrel{d_i}{\longrightarrow}
    \underline{\Z}\to 0 \big] 
  \right) = 
  \bigoplus_{j=1}^{2^{n-1}}
  \Z_{\gcd(d_1,\dots,d_n)}
\end{equation}
\begin{proof}
  Induction: It is true for $n=1$ and
  \begin{multline}
    H_\ast\left( \bigotimes_{i=1}^{n+1}
      \big[ 0\to \Z \stackrel{d_i}{\longrightarrow}
      \underline{\Z}\to 0 \big] 
    \right) 
    = \\ =
    \bigoplus_{j=1}^{2^{n-1}} 
    \Big( \Z_{\gcd(d_1,\dots,d_n)}\otimes \Z_{d_{n+1}} \Big)
    \oplus 
    \bigoplus_{j=1}^{2^{n-1}} 
    \Tor_\Z \left(
      \Z_{\gcd(d_1,\dots,d_n)}
      , \Z_{d_{n+1}} \right) 
    = \\ =
    \left(
      \bigoplus_{j=1}^{2^{n-1}} \Z_{\gcd(d_1,\dots,d_{n+1})}
    \right)
    \oplus 
    \left(
      \bigoplus_{j=1}^{2^{n-1}} \Z_{\gcd(d_1,\dots,d_{n+1})}
    \right)
  \end{multline}
\end{proof}
Finally note that $\gcd(d_1,\dots,d_{n})$ does not depend on the
choice of generators since it is the generator of the image
$\dim(I_k)\subset \Z$ under the dimension homomorphism $\dim:RG\to
\Z$. Explicitly computing $\gcd(d_1,\dots,d_{n})$ in any special case is
straightforward but tedious. Fortunately general expressions were
determined in~\cite{Bouwknegt}, and I will use their results (although
they do not prove every formula).

Combining eqns.~\eqref{eq:Kactionmap},~\eqref{eq:tKtorResult}
and~\eqref{eq:torresult} we determined $\tK^\ast(G)$. One can also
determine the contributions to $\tK^{0,1}$ separately by
keeping track of even/odd degree terms in $\Tor_{RG}^\ast$.  This is
in principle clear but gets somewhat messy to write down, so I avoided
it so far. The final result is that
\begin{equation}
  \label{eq:result}
  \tK^{\dim\,G}(G) = \big( \Z_x \big)^{r_0}
  \,,\quad 
  \tK^{1+\dim\,G}(G) = \big( \Z_x \big)^{r_1}
  \,,\quad 
  \tK^\ast(G)   = \big( \Z_x \big)^{r}
\end{equation}
where
\begin{equation}
  \label{eq:resultdefs}
  x \eqdef \frac{t_G}{ \gcd\Big(t_G, y \Big)}    
  \,,\qquad
  t_G = k + \check{h}(G)
\end{equation}
with all numerical coefficients determined by table~\ref{tab:result}.
\begin{table}[htb]
  \centering
  \begin{tabular}{c|ccccc}
    $G$ & $y$ & $r_0$ & $r_1$ & $r$ & $\check{h}$ 
     \\[1ex] \hline \strut
    $A_1=SU(2)$ & $1$ 
    & $1$ & $0$ & $1$ & $2$ \\[2ex]
%    $A_n$, $n\geq 2$ & 
    \parbox{2cm}{\raggedright
      $A_n$, $n\geq 2$ \\ $[SU(n+1)]$
    } & 
    $\lcm(1,2,\dots,n)$ 
    & $2^{n-2}$ & $2^{n-2}$ & $2^{n-1}$ & $n+1$ \\[2ex]
    $B_n$, $n\geq 2$ & 
    $\lcm(1,2,\dots,2n-1)$ 
    & $2^{n-2}$ & $2^{n-2}$ & $2^{n-1}$ & $2n-1$\\[2ex]
    $C_n$, $n\geq 3$ & 
%    $\lcm(1,2,\dots,n,1,3,5,\dots,2n-1)$ 
    \parbox{4cm}{\raggedright
      $\lcm(1,2,\dots,n,$  \\ \hfill
      $,1,3,5,\dots,2n-1)$ 
    }
    & $2^{n-2}$ & $2^{n-2}$ & $2^{n-1}$ & $n+1$ \\[2ex]
    $D_n$, $n\geq 4$ & 
    $\lcm(1,2,\dots,2n-3)$ 
    & $2^{n-2}$ & $2^{n-2}$ & $2^{n-1}$ & $2n-2$ \\[2ex]
    $G_2$  & $\lcm(1,2,\dots,5)=60$ 
    & $1$ & $1$ & $2$ & $4$ \\[2ex]
    $F_4$  & $\lcm(1,2,\dots,11)=27720$ 
    & $2^2$ & $2^2$ & $2^3$ & $9$ \\[2ex]
    $E_6$  & $\lcm(1,2,\dots,11)=27720$ 
    & $2^4$ & $2^4$ & $2^5$ & $12$ \\[2ex]
    $E_7$  & 
%    $\lcm(1,2,\dots,17)=12252240$ 
    \parbox{4cm}{\raggedright
      $\lcm(1,2,\dots,17)=$ \\ \hfill
      $=12252240$ 
    }
    & $2^5$ & $2^5$ & $2^6$ & $18$ \\[2ex]
    $E_8$  & 
%    $\lcm(1,2,\dots,29)=2329089562800$ 
    \parbox{4cm}{\raggedright      
      $\lcm(1,2,\dots,29)=$ \\ \hfill
      $=2329089562800$ 
    }
    & $2^6$ & $2^6$ & $2^7$ & $30$ \\
  \end{tabular}
  \caption{Coefficients determining the twisted \Kgroups{}, 
    see eq.~\eqref{eq:result}}
  \label{tab:result}
\end{table}

\section{There are no Exceptions}
\label{sec:exceptions}

For certain Lie groups and low levels ($k=1$ or $2$) it was noted
in~\cite{Bouwknegt} that some of the fundamental representations are
no longer in the Verlinde algebra. So they proposed that the
corresponding generator should be removed, and the CFT rule for
computing the \Dbrane{} charges should be applied to this presentation
of the algebra. Of course this then depends on the explicit
presentation of the Verlinde algebra, i.e. the choice of generators
and relations.

However (although far from obvious) in the \Ktheory{} computation
above there is no ambiguity since $\Tor$ is independent of the chosen
resolution. There is only a technical problem of computing the
$RG$--module action of the Verlinde Algebra in the exceptional cases
of~\cite{Bouwknegt}, which I want to comment on.

So, for example, consider $G_2$ at level $k=1$ as in~\cite{Bouwknegt}.
The Verlinde algebra is
\begin{subequations}
\begin{align}
  \label{eq:G2pres1}
  V_k(G_2) =&\,
  \Z[x_1, x_2] \Big/ \!
  \left< x_1,~ x_2^2-x_1-x_2-1 \right>
  \\  =&\,
  \label{eq:G2pres2}
  \Z[x_2] \Big/ \!
  \left< x_2^2-x_2-1 \right>    
\end{align}
\end{subequations}
where $x_1$ and $x_2$ are the two fundamental representations of
dimensions $14$ and $7$.

The generators for the relations in eq.~\eqref{eq:G2pres1} are a
regular sequence in $RG_2$ and we obviously recover the result of
table~\ref{tab:result} if we compute $\tK(G_2)$ that way.

But we are certainly allowed to use the simpler presentation
eq.~\eqref{eq:G2pres2} to compute $E_2$ in the \Kunneth{} spectral
sequence. We find that
\begin{multline}
  E_2^{-p,\ast} 
  =\Tor_{RG_2}\left(  
    \Z[x_2] \Big/ \! \left< x_2^2-x_2-1 \right>    
    ,~\Z
  \right) = 
  \\ =\,
  H_{-p}\Big(
  \xymatrix{
    0 \ar[r] &
    \Z[x_2] \otimes_{RG_2} \Z \ar[r]^{41} &
    \underline{\Z[x_2] \otimes_{RG_2} \Z} \ar[r] &
    0
  }
  \Big)
\end{multline}
But now $\Z[x_2] \otimes_{RG_2} \Z \simeq \Z_7$ since 
\begin{equation}
  7 \big( q \otimes n\big) = 
  q \otimes (7 n) = (x_1 q) \otimes n = 0 \otimes n = 0
\end{equation}
Since $41$ is invertible in $\Z_7$ we get
\begin{equation}
  E_2^{-p,\ast} 
  =
  H_{-p}\Big(
  \xymatrix{
    0 \ar[r] &
    \Z_7 \ar[r]^{41} &
    \underline{\Z_7} \ar[r] &
    0
  } \Big)
  = 0  
  \quad \forall p
\end{equation}
and thus we reproduce the result of table~\ref{tab:result}. 

This fits together nicely with the physical RG flow picture. Recall
that the flows that lead to eq.~\eqref{eq:RG} come from perturbations 
\begin{equation}
  S ~\to~ S+
  \Tr \textbf{P} \exp\left( \oint_{\partial \Sigma} J\right)
\end{equation}
by the holonomy of the spin current $J=\Lambda_\mu J^\mu$ in some
representation. Here you may choose any representation of $G$,
regardless of what its image in the Verlinde algebra would be. We have
to take all possible flows into account when we determine the
\Dbrane{} charge group, i.e. we would get the wrong answer if we
artificially restricted ourselves to representations which would be
nonzero in the Verlinde algebra.

\section{An algebra of BPS states}
\label{sect:multiplication}

It remains to show that the \Kunneth{} spectral sequence for
$\tK^\ast_{G}\big(G_\text{Ad}\times G_\text{L}\big)$ does actually
collapse at the $2^\text{nd}$ term, and that we can solve the
extension problems. Both can be dealt with by considering
multiplicative structures\footnote{Of course this has nothing to do
  with the algebra of BPS states as in~\cite{HarveyMooreBPS}.}.

Of course twisted \Ktheory{} does not come with a multiplication: The
twist class adds when you tensor twisted bundles. Instead I will work
in K--homology and use the fact that all spaces we consider are
actually groups, so they come with multiplication maps $\mu:X\times
X\to X$. Push forward induces the Pontryagin product
\begin{equation}
  \label{eq:pontryagin}
  \tK^G_\ast(X) \otimes_{RG} \tK^G_\ast(X) 
  \longrightarrow
  \tK^G_\ast(X\times X)
  \stackrel{\mu_\ast}{\longrightarrow}
  \tK^G_\ast(X)
\end{equation}
For $\tK^G_\ast(G_\text{Ad})$ it was noted in~\cite{FreedICM} that the
Pontryagin product is simply the fusion product in the Verlinde
algebra. Since it is easy to identify the Pontryagin product for the
trivial group $\ptset$ we see that
\begin{equation}
  \label{eq:pontryaginproductstructures}
  \begin{split}
    \tK^G_\ast(G_\text{Ad}) =& \, RG/I_k \\
    K^G_\ast(G_\text{L}) =& \,
    K_\ast \big( \ptset \big) = \Z \\
    K^G_\ast \big( \ptset \big) =& \, RG
  \end{split}
\end{equation}
as rings.

There is a dual version of the \Kunneth{} spectral sequence (the bar
spectral sequence of~\cite{RavenelWilson}) with
\begin{equation}
  \label{eq:bss}
  E^2_{p,\ast} = \Tor^{RG}_{p}
  \Big( \tK^G_\ast(X),~ \tK^G_\ast(Y) \Big)
  \quad \Rightarrow ~ \tK^G_\ast\big(X\times Y\big)
\end{equation}
so as before we use the trick with the action map of
section~\ref{sec:actionmap} to get a spectral sequence
\begin{equation}
  \label{eq:bsstK}
  E^2_{p,\ast} = \Tor^{RG}_{p}
  \Big( RG/I_k,~ \Z \Big)
  \quad \Rightarrow ~ 
  \tK^G_\ast\big(G_\text{Ad}\times G_\text{L}\big)
  = \tK_\ast(G)
\end{equation}
The advantage here is that it is a spectral sequence of
$RG$-algebras. 

Lets have a closer look at the $E^2$ term. The $\Tor(\cdot, \cdot)$ of
two algebras is again an algebra: Again I will use the Koszul
resolution of the Verlinde algebra (see section~\ref{sec:koszul}),
which can be described as the exterior algebra on $n=\rank(G)$
generators of degree $1$.
\begin{equation}
  \mathcal{K}(y_1, \dots, y_n) = 
  \Lambda_{RG}(\eta_1, \dots, \eta_n)
\end{equation}
The Verlinde algebra is then the homology of this exterior algebra
with respect to the differential $d(\eta_i)=y_i$. Applying $-\otimes
\Z$ and taking the homology with respect to $d(\eta_i)=\dim_\C(y_i)$
we determine the algebra structure of the $E^2$ term as
\begin{equation}
  \label{eq:toralg}
  \Tor^{RG}_{p}
  \Big( RG/I_k,~ \Z \Big)
  = \Lambda_{\Z_x}(\eta_1, \dots, \eta_n) 
\end{equation}
So the $E^2$ term looks like this
\begin{equation}
  \label{eq:E2differential}
  E^2_{p,\ast} = 
  \xymatrix@R=0mm@!C=20mm{
    \ar[]+/d 0.6cm/+/l 0.6cm/;[rrr]+/d 0.6cm/+/l 0.6cm/+/r 2cm/
    0 
    & \displaystyle
    \Z_x 
    & \displaystyle
    \bigoplus_{i=1..n} \!\! \eta_i \Z_x 
    \ar@/_8mm/[ll]_{d_2}
    & \displaystyle
    \bigoplus_{1\leq i<j\leq n} \!\!\!\!\! \eta_i \wedge \eta_j \Z_x
    \ar@/_8mm/[ll]_{d_2}
    \\
      {\scriptstyle p=-1} 
    & {\scriptstyle p=0} 
    & {\scriptstyle p=1} 
    & {\scriptstyle p=2} 
  }
\end{equation}
and $d_2$ obviously vanishes on $E^2_{1,\ast}$. Since all the algebra
generators are at degree $1$, the differential $d_2$ vanishes
identically. By the same reason all higher differentials are zero, and
$E^2=E^\infty$.

It remains to see that $\tK_\ast(G)$ is only $x$--torsion, in
principle there could be nontrivial extensions $0\to \Z_x\to
\Z_{x^2}\to \Z_x\to 0$ when we try to recover the homology from the
associated graded groups. The $E^\infty$ term of the spectral sequence
are the successive quotients $E^\infty_{p,\ast} = F^p/F^{p-1}$ of a
filtration
\begin{equation}
  0 \subset \cdots \subset 
  F^{p-1} \subset F^p \subset F^{p+1}
  \subset \cdots \subset 
  \tK_\ast(G)\,.
\end{equation}
The first quotient is $\Z_x = E^\infty_{0,\ast} = F^0/0 = F^0$, and
from the multiplicative structure of the exterior algebra we see that
this is a ring with unit. Since $1\in \Z_x \subset \tK_\ast(G)$ acts
as the identity on the successive quotients one can show by a simple
induction that it is actually the identity in the bigger ring
$\tK_\ast(G)$. But $1$ is $x$--torsion, and thus everything:
\begin{equation}
  x \kappa = (x\cdot 1)\kappa = 0\kappa = 0 
  \quad \forall \kappa \in \tK_\ast(G)
\end{equation}
By similar arguments one can show that
$\tK_\ast(G)=\Tor^{RG}_{\ast} \big( RG/I_k,~ \Z \big)$ as a ring.

\section{Level--Rank Nonduality}
\label{sec:levelrank}

For WZW models there are various level--rank dualities,
see~\cite{FuchsReview}. Those are distinct WZW models whose fusion
rings happens to be the same, for example $B_2$ at level $k=1$ and
$E_8$ at level $k=2$. The corresponding Verlinde algebras are
\begin{subequations}
\begin{align}
  \label{eq:verlindeB2}
  \big(B_2\big)_1 :&\quad
  RB_2/V_1 = \Z[x_1,x_2] \Big/ \!\!
  \left<
    x_2^2 =  1
    ,~
    x_2  x_1 =  x_1
    ,~
    x_1^2 =  1  + x_2
  \right>
  \\ 
  \notag
  \big(E_8\big)_2 :&\quad
  RE_8/V_2 = \Z[y_1,\dots,y_8] \Big/
  \\
  \label{eq:verlindeE8}
  &\quad \hphantom{RE_8/V_2 = \Z} 
  \Big/ \!\!
  \left<
    y_8^2 =  1
    ,~
    y_8  y_1 =  y_1
    ,~
    y_1^2 =  1  + y_8
    ,~
    y_2,\dots,y_7
  \right>
\end{align}
\end{subequations}
where the nontrivial generators
\begin{equation}
  \dim_\C(x_1)=4 \quad
  \dim_\C(x_2)=5 \quad
  \dim_\C(y_1)=248 \quad
  \dim_\C(y_8)=3875 \quad  
\end{equation}
correspond to the outermost nodes in the Dynkin diagram.

Now the two Verlinde algebras
eq.~\eqref{eq:verlindeB2},~\eqref{eq:verlindeE8} are obviously
isomorphic algebras. These algebras appear in the computation of the
twisted \Ktheory{} for
\begin{equation}
  \label{eq:levelrankgroups}
  B_2~\text{with}~t_{B_2} = 1+\check{h}(B_2) = 4
  \,,\qquad
  E_8~\text{with}~t_{E_8} = 2+\check{h}(E_8) = 32
  \,.
\end{equation}
From table~\ref{tab:result} we readily find
\begin{equation}
  \tK\big(B_2\big) = \Z_2 
  ~\not=~
  \Z_2^{128} = \tK\big(E_8\big)  
\end{equation}
Of course this should not be too much of a surprise, as the
computation of the twisted \Ktheory{} depends on the $RB_2\big/RE_8$
module structure of the Verlinde algebra. There is no reason why
level--rank duality should hold at the \Ktheory{} level.

This is in contrast to the supersymmetric Kazama--Suzuki
models~\cite{KazamaSuzuki} where level--rank duality is believed to be
an exact duality.

\section*{Acknowledgments}
\label{sec:acknowledgements}

I especially would like to thank Sakura Sch\"afer--Nameki for sharing
the draft of her upcoming paper with me. Furthemore it is a pleasure
to acknowledge useful discussions with Stefan Fredenhagen and Andreas
Szenes.

\bibliographystyle{JHEP}
\renewcommand{\refname}{Bibliography}
\addcontentsline{toc}{section}{Bibliography}
\bibliography{twistedktheory}

\end{document}